# Can students apply the concept of "which-path" information learned in the context of Mach Zehnder Interferometer to the double-slit experiment?

Alexandru Maries[1], Ryan Sayer[2] and Chandralekha Singh[3]

[1]*Department of Physics, University of Cincinnati, Cincinnati, OH 45221, USA*
[2] *Department of Physics, Bemidji State University, Bemidji, MN 56601 USA*
[3]*Department of Physics and Astronomy, University of Pittsburgh, Pittsburgh, PA 15260, USA*

**Abstract:** The ability to use concepts learned in one context to solve problems in a different context (i.e., transfer of learning) is often one of the major goals of physics instruction. After all, the ability to transfer learning from one context to another is a prerequisite to recognizing the applicability of compact physical laws to a multitude of contexts, and is a hallmark of expertise in physics. The majority of the studies on transfer of learning in physics have focused on introductory students and it has often been found that introductory students have difficulty effectively transferring their learning and applying what they learned in one situation to another situation. However, compared to advanced students, introductory students have significantly less prior relevant knowledge and skills which may be crucial for promoting effective transfer. Here, we examine upper-level undergraduate and graduate students' ability to transfer their learning about the concept of "which-path" information and its relation to whether or not interference is observed from the context of the Mach Zehnder Interferometer (MZI) to a new context of the double-slit experiment (DSE). Students worked through a tutorial on the MZI in which they learned to use the concept of "which-path" information to reason about interference of single photons when polarizers with various orientations are placed in one or both paths of the MZI. After working on the MZI tutorial, students were asked similar questions in the isomorphic context of the DSE before any instruction about the DSE and we examined the extent to which transfer of learning about "which-path" information occurred from the MZI to the DSE context. We find evidence suggesting that positive transfer of learning from the MZI to the DSE occurred despite the lack of an instructional intervention designed to help students recognize the similarity between the two contexts. The effectiveness of the MZI tutorial in promoting positive transfer of learning from one context to another in quantum mechanics sheds light on the expertise of the advanced physics students. Instructors of advanced physics courses can take advantage of the findings presented here which shows the difference between introductory and advanced physics students' expertise and ability to transfer from one context to another.

## I. INTRODUCTION

Physics is a domain in which a large number of phenomena are described by relatively few physical laws (e.g., conservation of energy, conservation of momentum, Maxwell's equations). Physics experts are very good at recognizing the applicability of physical laws to a variety of contexts, while novices (i.e., students learning those concepts) often show significant difficulty applying a physical law they have learned in one context to another [1,2]. In order for students to become experts, they must learn to recognize the usefulness of physics concepts learned in one context to other contexts, i.e., develop the ability to transfer their learning across contexts. In fact, the way students are typically assessed in exams requires them to demonstrate that they can transfer learning to solve novel problems in different contexts from the ones discussed in class or homework. Therefore, it is beneficial for instructors to learn about 1) the extent to which students can transfer their learning from one context to another and 2) effective approaches to help promote transfer of learning in the classroom. Prior studies suggest that transferring learning from one context to another context can be difficult especially if the "source" (from which transfer is intended) and the "target" (to which transfer is intended) do not share surface features [3]. This difficulty arises because knowledge is encoded in memory with the context in which it was learned and solving the source problem does not automatically manifest its "deep" similarity with the target problem [3].

Transfer of learning from one context to another requires that students engage in problem solving in a deep meaningful way and use it as an opportunity for extending and organizing their knowledge structure. It is therefore not surprising that developing expertise in physics problem solving and ability to transfer learning from one context to another constitutes a major goal of most physics courses [4]. Many studies on transfer of learning have focused on introductory students [5-7] and have found that students have significant difficulty transferring learning from one context to another, e.g., recognizing that the same physics principles they used to solve one problem are required to solve a similar or isomorphic problem involving the same underlying principles. Many introductory students are unable to transfer their learning from one problem to another, even when the two problems are posed one after the other as part of the same quiz [7].

In contrast to the many studies on transfer of learning in introductory physics, there is a significant dearth of literature on transfer of learning in advanced physics. Advanced physics students are further from the "novice" state than introductory physics students, and it is possible that throughout their undergraduate coursework, they have developed certain problem solving and reasoning skills [8] which may facilitate positive transfer. For example, Schraw [8] discusses skills which some students develop which can help them control their learning, and one of the skills is monitoring, or the on-line awareness of comprehension and task performance. In other words, an advanced physics student may have a better sense of his or her own comprehension of a particular concept, which can lead to more appropriate practice to evaluate one's own understanding compared to an introductory student. An advanced student may engage in deeper learning than an introductory student, and since recognizing the deep structure or underlying physics principles in a particular problem is a pre-requisite to transfer of learning to a different situation, advanced students may be more likely to effectively transfer their learning.

This study was focused on student learning of fundamental quantum mechanics concepts such as the wave-particle duality, single photon interference, and quantum eraser [9-18] in the context of the Mach-Zehnder Interferometer, or MZI [11,19]. The goal was to use isomorphic problems [20] with different surface features to investigate the extent to which students were able to transfer their learning in advanced quantum mechanics from the context of the MZI to the context of the double slit experiment, or DSE [21-23] without an explicit intervention designed to help them make the connection between the different contexts. Student difficulties with quantum mechanics as well as effective approaches to improve student learning in quantum mechanics have been investigated, e.g., see [24-66]. We developed an interactive tutorial on the Mach-Zehnder Interferometer (MZI) [19] which guides students to use the concept of "which-path" information (WPI) [17,67] to reason about the interference of single photons in the MZI with polarizers in various orientations placed in one or both paths of the MZI (we refer to these questions as "MZI polarizer questions"). A fundamental concept in quantum mechanics is interference of single particles, e.g., photons, which can be understood using WPI reasoning as was nicely illustrated in Ref. [42] by Muller and Wiesner to teach the basic principles of quantum mechanics (we provide an operational definition of WPI and how it can be used to reason about single-photon interference in the MZI and DSE contexts in the next section). We should note that helping students reason about interference in the MZI polarizer questions was a major goal of the MZI tutorial. Other goals included being able to calculate the probability of a certain detector to register a photon, what happens to measurement outcomes when a detector is placed in one path or the other, and the effect on the measurement outcomes when one of the beam splitters is removed. After working on the MZI tutorial which helps students learn how to use WPI to reason about interference in the MZI polarizer questions, students were given a pretest on the Double-Slit Experiment (DSE) which included questions with polarizers placed in front of one or both slits of the DSE. We refer to these questions as the "DSE polarizer questions" (described in detail in Section III) because they are isomorphic (similar in underlying physics principles) to the MZI polarizer questions. A control group of graduate students did not work on the MZI tutorial before answering the DSE polarizer questions. We investigated the extent to which students who worked on the MZI tutorial (and thus had the opportunity to learn about WPI) transferred their learning of WPI to answer the DSE polarizer questions. We note that the MZI tutorial is available online.

We begin by describing the isomorphism between the MZI and the DSE contexts with regard to WPI before describing the investigation. We then present the results, discuss some possible reasons for the observed transfer of learning, and finally summarize the findings.

## II. ISOMORPHISM BETWEEN MZI AND DSE

Before recognizing the isomorphism between the MZI and the DSE, one must first understand how the concept of WPI can be used to reason about interference in each experiment. The concept of WPI at a detector may be useful when the state of the system is a superposition of two different spatial path states (e.g., in the case of MZI or DSE with single photons). In general, when both components of the path state can reach a detector, then WPI is unknown. On the other hand, when only one component of the path state can reach a detector, then we have complete which path information, or WPI is known.

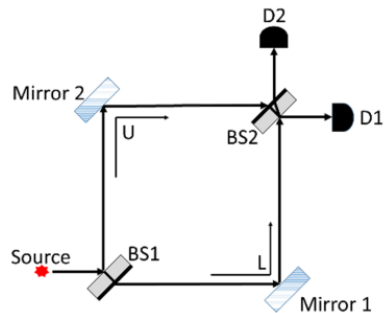

**Fig. 1**. Basic MZI setup.

We first consider the most basic MZI setup shown in Fig. 1. BS1 and BS2 are beam splitters; BS1 is oriented such that it puts the single photon emitted from the source into an equal superposition of the U and L path states shown (which we represent as $|U\rangle$ and $|L\rangle$, respectively), mirrors are included for proper alignment, and BS2 ensures that the components of the single photon state from both the U and L paths can reach each (photo) detector D1 and D2 after BS2 so that constructive or destructive interference (or anything in between) can be observed at the two detectors D1 and D2 in Fig. 1 (depending on the path length difference between the U and L paths). If an *additional* detector is placed anywhere in the lower path L between BS1 and BS2, after encountering the detector, the superposition of the U and L path states of a photon collapses and if the photon does not get absorbed by the detector, the state of the photon inside the MZI is the upper path state $|U\rangle$. Conversely, if an additional detector is placed in the upper path U, after encountering the detector, if the photon is not absorbed by that detector, the state of the photon inside the MZI collapses to the lower path state $|L\rangle$. In these situations (additional detector in the U or L path of the MZI), if a photon arrives at the detector D1 or D2 after BS2, we have WPI because only the component of the photon state along the U or L path can reach a detector and no interference is observed at D1 or D2. If instead, no detector is placed in either of the U or L path of the MZI (as in Fig. 1), the state of a photon inside the MZI remains an equal superposition of the U and L path states, WPI is unknown (because both the $|U\rangle$ and $|L\rangle$ components of the photon state can reach the detectors) and therefore interference is observed at D1 and D2.

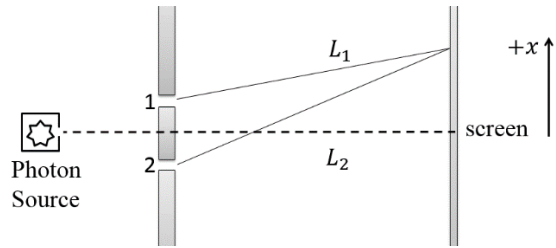

**Fig. 2**. Basic DSE setup with single photons.

Now consider the DSE setup shown in Fig. 2. If slit 2 is blocked, the state of a photon inside the DSE (after passing through the slits) collapses to $|\Psi_1\rangle$ and if slit 1 is blocked, the state of a photon collapses to $|\Psi_2\rangle$. If this photon arrives at the screen (the screen is the detection device in the DSE equivalent to detectors D1 and D2 in the MZI), we have WPI because only one component of the photon's path state (either $|\Psi_1\rangle$ or $|\Psi_2\rangle$) can reach the screen and therefore, no interference is observed. If neither slit is blocked, the state remains an equal superposition of $|\Psi_1\rangle$ and $|\Psi_2\rangle$. In other words, $|U\rangle$ and $|L\rangle$ in the MZI are analogous to $|\Psi_1\rangle$ and $|\Psi_2\rangle$ in the DSE. In the situations in which there is no detector in either path of the MZI and neither slit is blocked for the DSE, we do not have WPI and each photon interferes with itself.

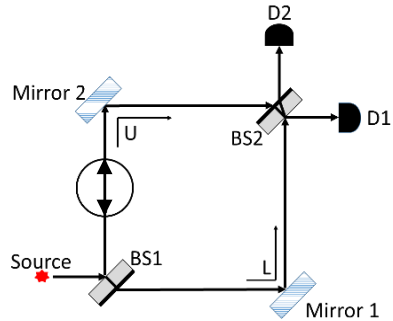

**Fig. 3**. MZI setup with a vertical polarizer placed in the upper path.

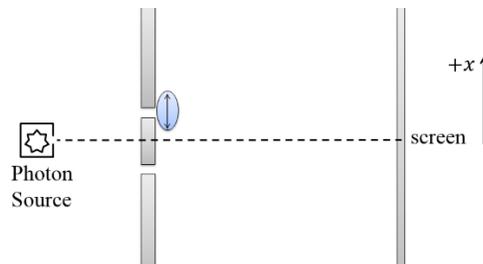

**Fig. 4.** DSE setup with a vertical polarizer placed after slit 1.

Now consider the situation shown in Fig. 3 in which we place a vertical polarizer in the upper path of the MZI and the source emits +45° polarized single photons. This situation is analogous to the situation shown in Fig. 4 in the DSE in which a vertical polarizer is placed after slit 1 (and the source emits +45° polarized single photons). We now have to use a four dimensional Hilbert space, two dimensions for path/slit states, $|U\rangle,|L\rangle/|\Psi_1\rangle,|\Psi_2\rangle$, and two dimensions for polarization states, for which the convenient basis for the situations described in Fig. 3 and Fig. 4 is $\{|V\rangle, |H\rangle\}$ (vertical, horizontal polarization states, respectively). If a vertical polarizer is placed in the upper path of the MZI, the $|U\rangle$ state will be associated with a vertical polarization state ($|U\rangle|V\rangle$) and the $|L\rangle$ state is still associated with both vertical and horizontal polarization states ($|L\rangle|V\rangle + |L\rangle|H\rangle$). In both experiments we assume that the detectors are sensitive to polarization (they are covered with polarizers with a particular orientation, e.g., vertical or horizontal), which means that the collapse of the photon state after it is measured by the detectors D1 or D2 provides information about the polarization of the photon. Therefore, in the situation depicted in Fig. 3, we have WPI for horizontally polarized photons arriving at D1 and D2 because the horizontal polarization is associated with the lower path state only –only the $|L\rangle$ component of the state of a horizontally polarized photon can reach the detectors. We do not have WPI for the vertically polarized photons because the vertical polarization is associated both with the upper and the lower path states – each detector can detect both the $|L\rangle$ and $|U\rangle$ components of the state of a vertically polarized photon. The fact that we have WPI for horizontally polarized photons and we do not have WPI for vertically polarized photons implies that the photons that arrive at the detectors in the $|V\rangle$ state interfere and those in the $|H\rangle$ state do not. In the DSE, the situation is analogous (Fig. 4): if a vertical polarizer is placed after slit 1, horizontally polarized photons arriving at the screen will not interfere, while vertically polarized photons arriving at the screen will show interference.

It is also possible to "erase" WPI by adding another polarizer (after BS2 in Fig. 3 for the MZI and right before the screen in Fig. 4 for the DSE) that is oriented in a different direction from the other polarizers. For example, for the MZI, suppose that the source emits photons with 45⁰ polarization and there is a vertical polarizer in the U path and a horizontal polarizer in the L path (equivalent to having a vertical polarizer after slit 1 and a horizontal polarizer after slit 2). In this situation, we would have WPI for all photons that arrive at detectors D1 and D2 because vertical and horizontal polarizations are associated with the upper and lower path, respectively. Thus, only one component of the path state for each polarization can reach the detectors and we would not observe interference for either polarization. However, if we place an additional polarizer oriented at 45⁰ between BS2 and D1, we will no longer have WPI for the photons arriving at D1 because now, both path states of the photon can reach the detector. This situation is referred to as a "quantum eraser" because the third polarizer "erases" WPI. In the case of the DSE, this would be equivalent to having a vertical polarizer after slit 1, a horizontal polarizer after slit 2, and an additional polarizer oriented at 45⁰ right before the screen.

Throughout this paper, we will refer to questions focusing on interference in these types of situations (single photons polarized at +45° emitted by the source and polarizers of various orientations placed in one or both paths of the MZI/in front of one or both slits of the DSE) as "MZI or DSE polarizer questions" depending on the context in which they are asked (MZI or DSE).

It is important to note that while the DSE and MZI contexts are isomorphic, the "surface" features of these two experiments are rather different. In the MZI, the paths are restricted and the photons arrive at point detectors D1 and D2, while in the DSE, the photons are delocalized in the space between the slits and the screen and can be detected anywhere on the extended screen. In addition, in the DSE, there is no explicit optical element corresponding to BS2 in the MZI which mixes the components of the photon state from the two paths (the screen itself does the mixing of the components of the single photon state from the two slits before the detection of the photon at the screen). These differences suggest that the surface features of these problems are quite different, which can make it challenging for novices to recognize the isomorphism. In order to recognize the isomorphism between the MZI and DSE, students must be able to reason about the deep features of the contexts and recognize the utility of the concept of WPI and its relation to whether or not interference will take place in both contexts. Thus, even if students fully understand the underlying physics principles in the MZI context, they may have difficulty recognizing how the same physics principles apply to the DSE.

### III. STUDY DESIGN AND FOCUS

The materials used in this study are the MZI pretest and posttest, the MZI tutorial and the DSE pretest. By working on the MZI tutorial, students learn how photo-detectors and optical elements such as beam-splitters in the path of the MZI with single photons affect the measurement outcomes. In addition, the MZI tutorial discusses setups in which polarizers of various orientations are placed in one or both paths and guides students to reason in terms of WPI to predict the outcome at the detectors. Thus, the MZI tutorial provides explicit help for answering the MZI polarizer questions which describe situations that are isomorphic to situations in the DSE polarizer questions. The development of the MZI tutorial included think-aloud interviews with both graduate and undergraduate students during which students worked on the tutorial while thinking out loud. Approximately 50 hours of individual student interviews were conducted, each interview lasting 2-4 hours. While students worked on the tutorial, they were not disturbed, and after they were finished, they were asked for clarification on points they had not made clear earlier while thinking out loud. In addition to the interviews, five physics faculty members were consulted several times during the development of the tutorial to ensure that they also found the wording of the questions unambiguous. Their feedback along with the student interviews was helpful in ensuring that the topics in the MZI tutorial were addressed appropriately and unambiguously.

The participants in this study were 46 undergraduate students enrolled in an upper-level quantum mechanics course and 59 physics graduate students enrolled in a mandatory semester-long TA professional development course which met for two hours each week. The 46 undergraduate students and 45 of the graduate students were part of the first cohort, and the other 14 graduate students were part of the second cohort. The undergraduates and graduate students in the first cohort took the MZI pretest, then worked on the MZI tutorial in small groups, and then took the MZI posttest. After completing all these three materials, they took the DSE pretest (which included the DSE polarizer questions, summarized below). We refer to these students in the first cohort as the "MZI→DSE cohort" to indicate that they completed the MZI tutorial before answering the DSE polarizer questions as part of the DSE pretest. The 14 graduate students in the second cohort did not work on the MZI tutorial before completing the DSE pretest. Therefore, we refer to these students as the "DSE only cohort". Information about how this relates to investigating transfer is provided in the next paragraphs which describe our research questions.

To investigate the extent to which students could transfer their learning from the context of the MZI to the context of the DSE, we used five questions which we refer to as "DSE polarizer questions" (asked as part of the DSE pretest). The situations for the questions are summarized as follows, and how they were used to investigate transfer of learning is clarified in our research questions:

"You perform a DSE in which photons that are polarized at +45° are sent one at a time towards the double slit. The wavelength of the photons is comparable to the slit width and the separation between the slits is more than twice the slit width. In all questions, assume that the same large number *N* of photons reaches the screen. In each situation, describe the pattern you expect to observe on the screen. Explain your reasoning."

Q1. Situation described above.
Q2. Vertical polarizer placed in front of one slit.
Q3. Vertical polarizer placed in front of each slit.
Q4. Vertical and horizontal polarizer placed in front of slits 1 and 2, respectively.
Q5. Vertical and horizontal polarizer placed in front of slits 1 and 2, respectively. Additionally, a polarizer which makes an angle of +45° with the horizontal is placed in between the slits and the screen.

These questions are analogous to questions students were asked in the context of the MZI. For example, the situation in Q2 above is analogous to the MZI setup shown in Fig. 3 in which a vertical polarizer is placed in one path. We note that questions 1 and 3 in the MZI context did not directly ask about the interference pattern, but rather asked about the fraction of photons that would be detected by D1 and D2, which are somewhat different questions although closely related to the presence/absence of interference). We also note that the DSE polarizer questions above were part of a larger pretest about the DSE which had other questions related to the DSE with single particles with mass, and we will not discuss these questions in detail here since they do not relate to the focus of this investigation, namely, transfer of learning. These questions are described in detail in Refs. [22] and [23].

If students in the MZI→DSE cohort transferred their learning appropriately, they should be more likely to make use of WPI reasoning and perform better on the DSE polarizer questions *after* working on the MZI tutorial compared to on the isomorphic MZI polarizer questions *before* working on the MZI tutorial. This is the focus of our first research question (broken in two parts for clarity), which includes only students from the MZI→DSE cohort.

**RQ1.**a. How does the percentage of students who used WPI reasoning *before* working on the MZI tutorial on MZI polarizer questions (on pretest) compare to the percentage of students who used WPI reasoning *after* working on the MZI tutorial on DSE polarizer questions?

**RQ1.**b. How does student performance on the MZI polarizer questions before working on the MZI tutorial compare to their performance on the isomorphic DSE polarizer questions after working on the MZI tutorial?

As noted, to further investigate the extent to which students can transfer their learning of WPI reasoning from the MZI to the DSE context, a second cohort of 14 graduate students (the DSE only cohort) did not work on the MZI tutorial before completing the DSE pretest. In other words, these students did not have the opportunity to learn about WPI reasoning before answering the DSE polarizer questions. We compared how often students in the DSE only cohort used WPI reasoning and how they performed on the DSE polarizer questions with the students in the MZI→DSE cohort (who had the opportunity to learn about WPI reasoning before answering these questions). We hypothesized that if some students are able to transfer learning about WPI from the MZI to the DSE context, when we compare the performance of the MZI→DSE cohort to the DSE only cohort on the DSE pretest questions, we would observe a difference on the DSE polarizer questions, but no difference on the other DSE questions. This is the focus of RQ2.a and RQ2.b below.

**RQ2.a.** What percentage of students who worked on the MZI tutorial (MZI→DSE cohort) used WPI reasoning on the DSE polarizer questions, and how does that compare to the students who did not work on the MZI tutorial (DSE only cohort)?

**RQ2.b.** How does the performance on the DSE polarizer and other DSE questions of students who worked on the MZI tutorial (MZI→DSE cohort) compare to that of students who did not work on the MZI tutorial (DSE only cohort)?

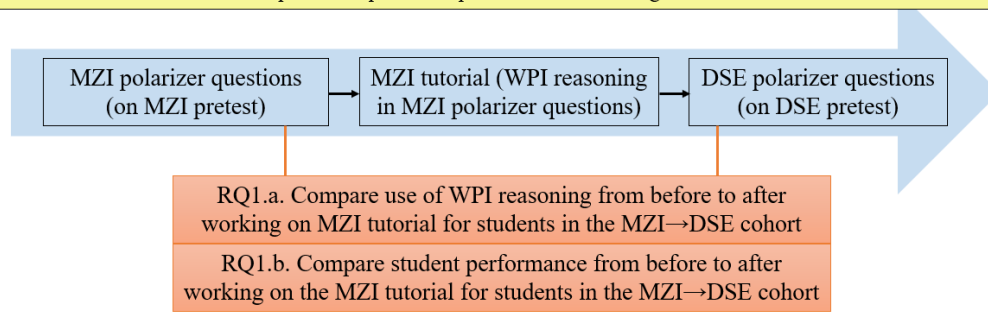

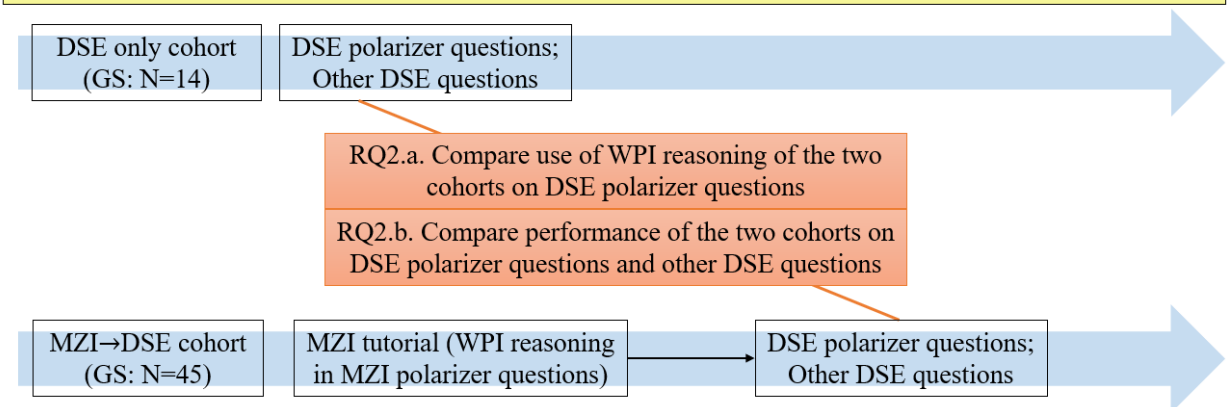

**Fig. 5.** Schematic description of the design used to investigate RQ1 and RQ2. The timeline showing the order in which students completed the materials is depicted by the blue arrows. In the figure "GS" refers to graduate students. Students included in answering RQ1 are the 45 graduate students and 46 undergraduate students from the MZI→DSE cohort, and the students included in answering RQ2 are the 45 graduate students from the MZI→DSE cohort and the 14 graduate students from the DSE only cohort.

To investigate RQ2.a, we compared the two cohorts of graduate students (MZI→DSE cohort with DSE only cohort) in terms of the percentage of students who used WPI reasoning on the DSE polarizer questions on the DSE pretest. To investigate RQ2.b, we conducted a one-way repeated measures ANOVA [68] to determine which interactions between condition (worked on MZI tutorial/did not work on MZI tutorial) and performance (DSE polarizer questions/other DSE questions) prove to be significant. If students are indeed transferring learning from the MZI tutorial to answer the DSE polarizer questions and are not just more likely to answer these questions correctly due to chance, the only significant interaction should be between the condition of working on the MZI tutorial and the DSE pretest polarizer question performance. In other words, we should find that the MZI tutorial has a significant positive effect on student performance on the DSE polarizer questions but little to no effect on the other DSE questions.

RQ1.a and RQ1.b along with the approaches used to investigate them are summarized in Fig. 5.

The summary of the rubric used to grade students' performance on the DSE polarizer questions is shown in Table 1. This rubric is designed to evaluate students' conceptual understanding of the effect of placing polarizers of various orientations in front of one or both slits in the DSE. For example, the third conceptual point (recognize that "which-path" information can be lost) is based on students' answers to the last two questions (orthogonal polarizers and quantum eraser). The same is true for the second conceptual point.

**TABLE 1.** Summary of the rubric used to grade students' performance on the DSE polarizer questions.

| | |
|---|---|
| Recognize that photons exhibit interference | +1, 0 |
| Recognize that <u>only</u>* when two polarizers are orthogonal and there is no "quantum eraser" – the interference pattern vanishes | +1, 0 |
| Recognize that "which-path" information can be lost | +1, 0 |
| Correctly interpret the effect of one polarizer on the interference pattern | +1, 0 |
| Correctly interpret the effect of two polarizers on the interference pattern (both questions) | +2, 1, 0 |

* If a student noted that interference pattern vanishes in more than 1 situation → 0 points

## IV.   RESULTS

*RQ1.a. How does the percentage of students who used WPI reasoning before working on the MZI tutorial on MZI polarizer questions (on pretest) compare to the percentage of students who used WPI reasoning after working on the MZI tutorial on DSE polarizer questions?*

For the MZI polarizer questions, we found that without any prior instruction related to WPI, only one graduate student (out of 45) used WPI reasoning in only one question. None of the 46 undergraduate students used this type of reasoning. Thus, it appears that neither graduate nor undergraduate students are aware of how to use this type of reasoning without prior instruction.

For the DSE polarizer questions 2-5, Table 2 shows the percentage of both undergraduate and graduate students from the MZI→DSE cohort who reasoned using WPI out of the students who provided any reasoning for their answers. We note that despite the fact that all questions explicitly asked for reasoning, some students did not provide any reasoning. However, these students may have answered the questions by reasoning about WPI while not explicitly writing it down. Table 2 shows that students from the MZI→DSE cohort often used WPI reasoning to answer the DSE polarizer questions, especially on the last two questions. In addition, out of all instances in which a graduate or undergraduate student used WPI reasoning to answer a question, he/she used it correctly 79% of the time, thus suggesting significant positive transfer of WPI reasoning learned in the MZI context to the DSE context.

**TABLE 2.** Percentage of undergraduate students (US) and graduate students (GS) from MZI→DSE cohort who used reasoning related to WPI out of those who provided reasoning on DSE polarizer questions 2-5 (Q2-Q5) in the pretest. All these students worked on the MZI tutorial prior to answering these questions. Combining graduate and undergraduate students, they used WPI reasoning correctly 79% of the time.

|    | Q2 | Q3 | Q4 | Q5 |
|----|----|----|----|----|
| US | 37 | 37 | 57 | 62 |
| GS | 33 | 20 | 60 | 44 |

*RQ1.b. How does student performance on the MZI polarizer questions before working on the MZI tutorial compare to their performance on the isomorphic DSE polarizer questions after working on the MZI tutorial?*

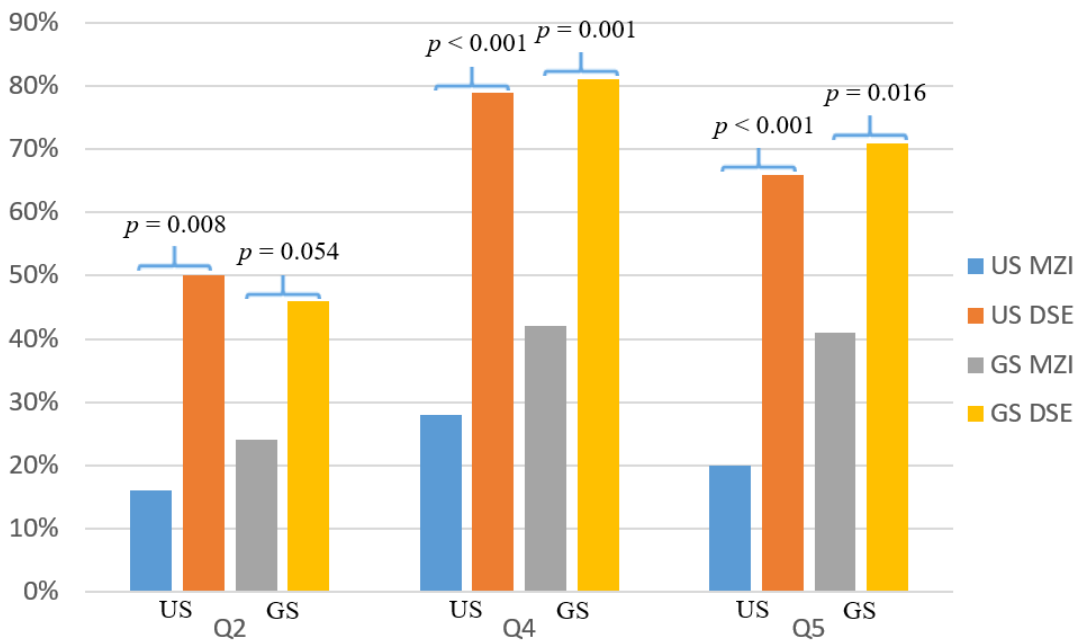

**Fig. 6.** Percentage of undergraduate (US) and graduate students (GS) in the MZI→DSE cohort who correctly answered 1) the polarizer questions in the context of the MZI on the MZI pretest, i.e., prior to any instruction related to WPI (US MZI and GS MZI), 2) the polarizer questions in the context of the DSE on the DSE pretest after receiving instruction on WPI from the MZI tutorial (US DSE, GS DSE), and $p$ values obtained from chi-square tests comparing these percentages. The average performance of undergraduate and graduate students on all three MZI polarizer questions was 21% and 36%, respectively, while for the DSE polarizer questions it was 65% and 66%, respectively.

Figure 6 shows the percentage of undergraduate and graduate students from the MZI→DSE cohort who answered MZI polarizer questions 2, 4, and 5 correctly on the MZI pretest (before any instruction related to WPI) and the percentage of undergraduate and graduate students who answered the isomorphic DSE polarizer questions correctly on the DSE pretest (after working on the MZI tutorial). We also investigated if there was a significant difference in the proportion of correct answers on each of the DSE questions with the proportion of correct answers on the isomorphic MZI questions from the MZI pretest and the $p$ values obtained by conducting chi-square tests [68] which are also shown in Fig. 6. Note that these are the same students (MZI→DSE cohort) who took the MZI pretest, then worked on the MZI tutorial, then took the DSE pretest. The *p* values show that with the exception of Q2 (the most challenging question for both student groups) both undergraduate and graduate students performed better in the DSE context than the MZI context. This suggests that some students were able to transfer their learning appropriately from the context of the MZI to answer the DSE polarizer questions. The difference in the percentage of students who answered the MZI polarizer questions correctly vs. the DSE polarizer questions ranges from 34% to 51% for the undergraduate students and 22% to 39% for the graduate students.

*RQ2.a. What percentage of students who worked on the MZI tutorial (MZI→DSE cohort) used WPI reasoning on the DSE polarizer questions, and how does that compare to the students who did not work on the MZI tutorial (DSE only cohort)?*

Table 2 shows the percentage of students from the MZI→DSE cohort who used WPI reasoning on the DSE polarizer questions. By comparison, out of the 14 graduate students from the DSE only cohort, only one graduate student used WPI reasoning in only one question. This provides further evidence that some of the students from the MZI→DSE cohort who had the opportunity to learn about WPI reasoning from the MZI tutorial transferred their learning to the different context of the DSE on their own without an explicit intervention designed to help them make the connection between these two contexts.

*RQ2.b. How does the performance on the DSE polarizer and other DSE questions of students who worked on the MZI tutorial (MZI→DSE cohort) compare to that of students who did not work on the MZI tutorial (DSE only cohort)?*

Table 3 shows the percentage of graduate students (depending on whether or not they worked on the MZI tutorial) who answered DSE polarizer questions Q1 through Q5 correctly. Although the numbers are small to perform meaningful statistics on each individual question, Table 3 suggests that students who had the opportunity to learn from the MZI tutorial were more likely to answer these questions correctly (meaningful statistics can however be performed on the aggregate data, i.e., overall performance on the DSE polarizer questions, see below). This is another indication of positive transfer from the MZI tutorial to the DSE polarizer questions.

Figure 7 shows the average performance of graduate students from the MZI→DSE and DSE only cohorts on the DSE polarizer questions (as graded using the rubric shown in Table 1), as well as their performance on the other DSE questions which were quite different from the DSE polarizer questions. A repeated measures ANOVA was carried out on these data which shows a statistically significant effect of engaging with the MZI tutorial on their performance on the polarizer questions on the DSE pretest ($p = 0.011$) [68]. None of the other interactions were significant suggesting that the MZI tutorial helped students only on the DSE polarizer questions (and not on other DSE questions). Furthermore, the effect size [68] for comparing the performance on the DSE polarizer questions of graduate students who worked on the MZI tutorial with the performance of those who did not was 0.831, thus suggesting a large effect of working on the MZI tutorial on these questions.

**TABLE 3.** Percentage of graduate students who answered DSE polarizer questions Q1 through Q5 correctly on the DSE pretest (depending on whether they worked or did not work on the MZI tutorial, i.e., whether they were in the MZI→DSE or DSE only cohort).

|  | Q1 | Q2 | Q3 | Q4 | Q5 |
|---|---|---|---|---|---|
| DSE only cohort (N=14) | 77% | 8% | 58% | 80% | 50% |
| MZI→DSE cohort (N=45) | 88% | 46% | 78% | 81% | 71% |

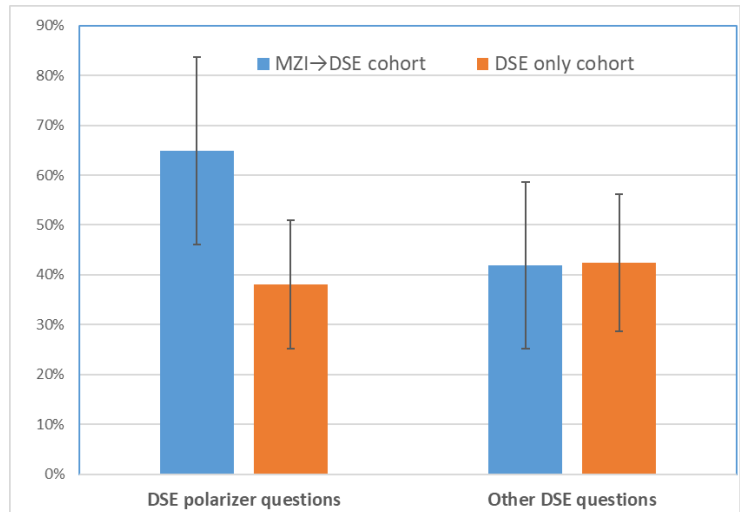

**Fig. 7.** Performance of the two graduate student cohorts (MZI→DSE cohort and DSE only cohort) on the DSE polarizer questions and on the other DSE questions on the DSE pretest. The error bars indicate standard deviations.

## V. POSSIBLE REASONS FOR TRANSFER FROM THE MZI TO THE DSE CONTEXT AND FUTURE RESEARCH

While it is difficult to identify the exact causes of the substantial transfer of learning from the MZI to the DSE context observed in this investigation, it is possible that upper-level undergraduate and graduate students have developed sufficient abstract reasoning skills which allow them to recognize the isomorphism between these situations and the usefulness of reasoning about WPI in both contexts. This is supported by the finding that many students provide WPI reasoning for the DSE polarizer questions on the pretest when they had only learned it in the MZI context compared to students who did not learn from the MZI tutorial. Reasoning in terms of WPI to answer questions related to interference of single photons was only discussed in the context of the MZI and students had to recognize the similarity between the MZI and DSE contexts in terms of underlying physics principles in order to answer the DSE polarizer questions correctly.

It is possible that both of these groups of students have developed sufficient skills to transfer their learning between these contexts which have different surface features. As discussed by Schraw [8], if the higher order thinking skills developed by a learner in a particular domain reach a certain threshold, these skills become more readily transferable to a new domain. Furthermore, if the domains share similar characteristics, transfer of these skills is more likely to occur. In this study, although both contexts are in physics, the contexts are sufficiently different. We can therefore interpret the transfer results found in this investigation to be partly due to many advanced students' ability to utilize higher order thinking skills to transfer their learning from one context to another.

Future study can investigate student reasoning more in-depth when answering the DSE polarizer questions both for students who worked through the MZI tutorial and those who have not. Think-aloud interviews can shed light on the extent to which students are recognizing the deep similarity of the two contexts and using the connection between the two to answer the DSE polarizer questions. If they are having difficulty recognizing the connection, guided questioning could shed light on what learning has transferred.

As mentioned earlier, prior investigations suggest that introductory students have significant difficulty transferring their learning and often, providing effective instruction to facilitate transfer is a daunting task. The study presented here shows promising results regarding transfer of learning for advanced students, and the extent to which transfer of learning is more common for advanced students compared to introductory students requires further research.

Another possible reason for the significant transfer found in this study is that while the isomorphism between the MZI and DSE is in underlying physics and the contexts are different, both use single photons and polarizers of various orientations placed after one or both slits or paths. This type of similarity may have prompted students to utilize analogous reasoning when answering the DSE polarizer questions. We note, however, that in the MZI posttest, student average scores were near the ceiling (~ 90%), while the averages on the DSE polarizer questions were around 70% for both undergraduates and graduate students after working on these concepts only via the MZI tutorial, implying that the transfer from the MZI to the DSE context is not perfect.

Finally, another possible reason for transfer is temporal proximity: after students worked on the MZI tutorial, they took the related posttest, and then the DSE polarizer questions were given. This proximity in timing may make it more likely for students to be able to discern the similarity between the two contexts and transfer their learning from the MZI context to the DSE context. However, as mentioned earlier, in introductory physics, even if two questions which require use of the same underlying physics principles are asked back to back as part of the same quiz, a majority of students may not discern the similarity between the questions and therefore answer them using different reasoning [7]. In future, one can investigate how transfer of learning is affected when there is a time lag, e.g., one month between the MZI tutorial and the DSE pretest.

## VI. SUMMARY

We investigated transfer of learning for advanced undergraduate and graduate students. We found evidence of significant transfer of learning from one context (MZI) to another (DSE). In particular, students took a pretest on the MZI after which they worked on the MZI tutorial which introduced students to the concept of WPI and guided them to use this concept to reason about whether or not interference is observed at the detectors in a particular MZI setup. When the DSE polarizer questions were administered afterwards, students who had worked on the MZI tutorial performed significantly better on the DSE polarizer questions (average of 65% and 66% for undergraduate and graduate students, respectively) than on the analogous MZI pretest questions (average of 21% and 36% for undergraduate and graduate students, respectively). Additionally, the graduate students, who worked on the MZI tutorial before answering the DSE polarizer questions on the DSE pretest, performed significantly better on these questions than the graduate students who did not work on the MZI tutorial. These two cohorts of graduate students showed identical performance on the other DSE questions which did not have analogous situations discussed in the MZI tutorial, thus suggesting that the improved performance on the DSE polarizer questions is likely due to transfer of learning rather than a difference in population. Another indication of transfer is that students often explicitly used reasoning learned in the context of the MZI to answer the DSE polarizer questions, e.g., they used reasoning related to WPI, and most students who used this type of reasoning did so correctly, indicating appropriate transfer from the MZI to the DSE context. Finally, students sometimes explicitly drew the parallel between the DSE and the MZI contexts themselves without any prompting.

We hypothesize that the observed transfer is partly due to the fact that advanced undergraduate and graduate students have developed sufficient abstract reasoning skills which allow them to recognize the isomorphism between the two contexts. This would in turn make it likely that they are able to apply analogical reasoning between the two contexts. Given that many students (both undergraduate and graduate) reasoned in terms of WPI to answer the DSE polarizer questions and that this reasoning was only discussed in the MZI context it appears that students may be recognizing underlying similarity of the physics in these situations. Additionally, the familiar context of the DSE and students' knowledge of polarizers may contribute to their ability to transfer their learning about WPI to answer the DSE polarizer questions. Informal discussions with graduate students who participated in this study suggested that they generally agreed with these two possible reasons. However, more research is required to elucidate the extent to which each of these plays a role in helping students discern the similarity between the two contexts. Other possible reasons include the close temporal proximity of the MZI tutorial to the DSE polarizer questions and the fact that both the MZI and DSE questions relate to single photons and polarizers placed after various paths/slits. However, as noted earlier, introductory students often have difficulty discerning the similarity between isomorphic problems even if they are placed back to back [7]. In addition, the differences between the setups suggest that the surface features of these problems are quite different which can make it challenging to recognize the isomorphism between the MZI and the DSE. Therefore, it is encouraging that advanced students have developed sufficient reasoning skills to be able to transfer their learning at least in the context discussed. Instructors of advanced physics courses can take advantage of the findings presented here which shows the difference between introductory and advanced physics students' expertise in physics.

## VII. ACKNOWLEDGEMENTS

We thank the National Science Foundation for awards PHY-1806691, Albert Huber for creating the simulation used in the MZI tutorial. We also thank F. Reif, R. D. Devaty and E. Marshman for helpful discussions.

## VIII. REFERENCES

1. W. Leonard, R. Dufresne, and J. Mestre, Using qualitative problem solving strategies to highlight the role of conceptual knowledge in solving problems, Am. J. Phys. **64**, 1495 (1996).
2. M. Gick and K. Holyoak, "The cognitive basis of knowledge transfer", in *Transfer of Learning: Contemporary Research and Applications* (Academic Press, 1987) pp. 9-46.
3. R. Benander, A Literature Summary for Research on the Transfer of Learning, EDU/EDPC (2018) 4
4. F. Reif, "Millikan Lecture 1994: Understanding and teaching important scientific thought processes", Am. J. Phys. **63**, 17 (1995).
5. L. Novick, "Analogical transfer, problem similarity and expertise", J. Exp. Psychol. Learn. **14** 510 (1988).
6. J. Lobato, "How design experiments can inform a rethinking of transfer and vice versa", Educ. Res. **32**, 17 (2003).

7. C. Singh, "Assessing Student Expertise in Introductory Physics with Isomorphic Problems. II. Effect of some potential factors on problem solving and transfer", Phys. Rev. ST PER **4** 010105 (2008).
8. G. Schraw, "Promoting general metacognitive awareness", Instructional Science **26** 113 (1998).
9. M. Arndt, O. Nairz, J. Vos-Andreae, C. Keller, G. van der Zouw, and Anton Zeilinger, "Wave-particle duality of $C_{60}$ molecules", Nature **401**, 680 (1999).
10. T. J. Herzog, P. G. Kwiat, H. Weinfurter, and A. Zeilinger, "Complementarity and the Quantum Eraser", Phys. Rev. Lett. **75**, 3034 (1995).
11. P. G. Kwiat, A. M. Steinberg, and R. Y. Chiao, "Observation of a ''quantum eraser'': A revival of coherence in a two-photon interference experiment", Phys. Rev. A **45**, 7729 (1992).
12. C. Bennett, Quantum cryptography using any two nonorthogonal states, Phys. Rev. Lett. **68**, 3121 (1992).
13. C. H. Bennett, G. Brassard, and N. D. Mermin, Quantum cryptography without Bell's theorem, Phys. Rev. Lett. **68** (5), 557 (1992).
14. C. H. Bennett, F. Bessette, G. Brassard, L. Salvail, and J. Smolin, Experimental quantum cryptography, Journal of Cryptology, **5** (1), 3 (1992).
15. D. H. McIntyre, *Quantum Mechanics* (Pearson Addison-Wesley, San Francisco, CA, 2012)
16. D. J. Griffiths, *Introduction to Quantum Mechanics, 2$^{nd}$ ed* (Prentice Hall, Upper Saddle River, NJ, 1995)
17. Christian Ferrari and Bernd Braunecker, "Entanglement, which-way measurements, and a quantum erasure", Am. J. Phys. **78**, 792 (2010).
18. J M. Ashby, P. D. Schwarz and M. Schlosshauer, "Delayed-choice quantum eraser for the undergraduate laboratory", Am. J. Phys. **84**, 95 (2016).
19. E. Marshman and C. Singh, "Interactive tutorial to improve student understanding of single photon experiments involving a Mach-Zehnder Interferometer", Eur. J. Phys **37**, 024001 (2017).
20. H. A. Simon and J. R. Hayes, "The understanding Process: Problem Isomorphs", Cog. Psych. **8**, 165 (1976).
21. W. Rueckner and J. Peidle, "Young's double-slit experiment with single photons and quantum eraser", Am. J. Phys. **81**, 951 (2013).
22. R. Sayer, A. Maries, and C. Singh, "A quantum interactive learning tutorial on the double-slit experiment to improve student understanding of quantum mechanics", Phys Rev PER **13**, 010123 (2017)
23. A. Maries, R. Sayer and C. Singh, Effectiveness of interactive tutorials in promoting which-path information reasoning in advanced quantum mechanics, Phys. Rev. PER **13**, 020115 (2017).
24. C. Singh, "Student Understanding of Quantum Mechanics", *Am. J. Phys.* **69**, 885 (2001).
25. E. Marshman and C. Singh, "Framework for understanding the patterns of student difficulties in quantum mechanics, Phys. Rev. ST Phys. Educ. Res. **11**, 020119 (2015).
26. C. Singh, "Assessing and improving student understanding of quantum mechanics", Proc. Physics Education Research Conf (2006). doi:10.1063/1.2177025.
27. G. Zhu and C. Singh, "Improving students' understanding of quantum mechanics via the Stern–Gerlach experiment", Am. J. Phys. **79**, 499 (2011)

28. G. Zhu and C. Singh, "Improving students' understanding of quantum measurement: I. Investigation of difficulties", Phys. Rev. ST Phys. Educ. Res. **8**, 010117 (2012).
29. G. Zhu and C. Singh, "Improving students' understanding of quantum measurement: II. Development of research-based learning tools", *Phys. Rev. ST Phys. Educ. Res.* **8**, 010118 (2012).
30. G. Zhu and C. Singh, "Surveying students' understanding of quantum mechanics in one spatial dimension", Am. J. Phys. **80**, 252 (2012).
31. S. Lin and C. Singh, "Categorization of quantum mechanics problems by professors and students", Eur. J. Phys. **31**, 57 (2010).
32. E. Marshman and C. Singh, "Investigating and improving student understanding of the probability distributions for measuring physical observables in quantum mechanics", Eur. J. Phys. **38**, 025795 (2017).
33. S. Siddiqui and C. Singh, How diverse are physics instructors' attitudes and approaches to teaching undergraduate-level quantum mechanics?" Eur. J. Phys. **38**, 035703 (2017).
34. B. Brown and C. Singh, "Development and evaluation of a quantum interactive learning tutorial on Larmor precession of spin", Proc. Physics Education Research Conf. (2015) doi:10.1119/perc.2014.pr.008.
35. C. Singh, M. Belloni, and W. Christian, Improving Students' Understanding of Quantum Mechanics, Phys. Today **8**, 43 (2006).
36. C. Singh, "Student Understanding of Quantum Mechanics at the Beginning of Graduate Instruction", Am. J. Phys. **76**, 277 (2008).
37. C. Singh, "Interactive Learning Tutorials on Quantum Mechanics", Am. J. Phys. **76**, 400 (2008).
38. C. Singh and G. Zhu, Cognitive Issues in Learning Advanced Physics: An Example from Quantum Mechanics, Proc. Physics Education Research Conf. (2009). (http://doi.org/10.1063/1.3266755).
39. C. Singh, Student difficulties with quantum mechanics formalism *Proc. Phys. Educ. Res. Conf.* (2006) (http://doi.org/10.1063/1.2508723)
40. P. Jolly, D. Zollman, S. Rebello and A. Dimitrova, "Visualizing potential energy diagrams", Am. J. Phys. **66**, 57 (1998).
41. D. Zollman, N. S. Rebello, and K. Hogg, "Quantum mechanics for everyone: Hands-on activities integrated with technology", Am. J. Phys. **70**, 252 (2002).
42. R. Muller and H. Wiesner, "Teaching quantum mechanics on an introductory level", Am. J. Phys. **70**, 200 (2002).
43. M. Wittmann, R. Steinberg et al., "Investigating student understanding of quantum physics: Spontaneous models of conductivity", Am. J. Phys. **70**, 218 (2002).
44. G. Passante et al., "Examining student ideas about energy measurements on quantum states across undergraduate and graduate levels", Phys. Rev. ST PER **11**, 020111 (2015).
45. E. Gire and E. Price, "The structural features of algebraic quantum notations," Phys. Rev. ST PER **11**, 020109 (2015).
46. V. Dini and D. Hammer, "Case study of a successful learner's epistemological framings of quantum mechanics," Phys. Rev. Phys. Educ. Res. **13**, 010124 (2017).
47. C. Keebaugh et al., Improving student understanding of fine structure corrections to the energy spectrum of the hydrogen atom, Am. J. Phys. **87**, 594 (2019).
48. Keebaugh, E. Marshman and C. Singh, Improving student understanding of a system of identical particles with a fixed total energy, C. Am. J. Phys. **87**, 583 (2019).

49. E. Marshman and C. Singh, Validation and administration of a conceptual survey on the formalism and postulates of quantum mechanics, Phys. Rev. PER **15**, 020128 (2019).
50. A. Kohnle, I. Bozhinova, D. Browne, M. Everitt, A. Fomins, P. Kok, G. Kulaitis, M Prokopas, D. Raine, and E. Swinbank, "A new introductory quantum mechanics curriculum", Eur. J. Phys. **35**, 015001 (2014).
51. A. Kohnle, C. Baily, A. Campbell, N. Korolkova, and Mark J. Paetkau, "Enhancing student learning of two-level quantum systems with interactive simulations", Am. J. Phys. **83**, 560 (2015).
52. A. Kohnle, D. Cassettari, T. J. Edwards, C. Ferguson, A. D. Gillies, C. A. Hooley, N. Korolkova, J. Llama, and B. D. Sinclair, "A new multimedia resource for teaching quantum mechanics concepts", Am. J. Phys. **80**, 148 (2012).
53. D. Domert, C. Linder, and A. Ingerman, "Probability as a conceptual hurdle to understanding one-dimensional quantum scattering and tunneling", Eur. J. Phys. **26**, 47 (2004).
54. E. Galvez and C. Holbrow, "Interference with correlated photons: Five quantum mechanics experiments for undergraduates", Am. J. Phys. **73**, 127-140 (2005).
55. M. N. Beck and M. Beck, "Witnessing entanglement in an undergraduate laboratory", Am. J. Phys. **84**, 87 (2016).
56. D. Branning, S. Bhandari and M. Beck, "Low-cost coincidence-counting electronics for undergraduate quantum optics", Am. J. Phys. **77**, 667 (2009).
57. A. López-Incera and W. Dür, "Entangle me! A game to demonstrate the principles of quantum mechanics", Am. J. Phys. **87**, 95 (2015).
58. E. J. Galvez, "Resource Letter SPE-1: Single-Photon Experiments in the Undergraduate Laboratory", Am. J. Phys. **82**, 1018 (2014).
59. C. Singh and E. Marshman, "Review of student difficulties in upper-level quantum mechanics", Phys. Rev. ST PER **11**, 020117 (2015).
60. P. Justice, E. Marshman, and C. Singh, "Improving student understanding of quantum mechanics underlying the Stern-Gerlach experiment using a research-validated multiple-choice question sequence", Eur. J. Phys. **40**, 055702 (2019).
61. P. Justice, E. Marshman, and C. Singh, "Student understanding of Fermi energy, the Fermi-Dirac distribution and total electronic energy of a free electron gas", Eur. J. Phys. **41**, 015704 (2020).
62. E Marshman and C Singh, "Investigating and improving student understanding of the expectation values of observables in quantum mechanics", Eur. J. Phys. **38**, 045701 (2017).
63. E Marshman and C Singh, "Investigating and improving student understanding of quantum mechanics in the context of single photon interference", Phys. Rev. PER **13**, 010117 (2017).
64. E Marshman and C Singh, "Investigating and improving student understanding of quantum mechanical observables and their corresponding operators in Dirac notation", Eur. J. Phys. **39**, 015707 (2017).
65. C. Keebaugh, E. Marshman and C. Singh, "Investigating and addressing student difficulties with the corrections to the energies of the hydrogen atom for the strong and weak field Zeeman effect", Eur. J. Phys. **39**, 045701 (2018).
66. C. Keebaugh, E. Marshman and C. Singh, "Investigating and addressing student difficulties with a good basis for finding perturbative corrections in the context of degenerate perturbation theory", Eur. J. Phys. **39**, 055701 (2018).

67. J. Wheeler 1979 The "Past" and the "Delayed-Choice" Double-Slit Experiment *Mathematical Foundations of Quantum Theory* (Academic Press) pp. 9-48.
68. Glass G V and Hopkins K D 1996 *Statistical Methods in Education & Psychology* (Allyn & Bacon).